\documentclass{article}
\usepackage{spconf,amsmath,graphicx}
\usepackage{subfigure}
\usepackage{setspace}
\usepackage{booktabs} 
\usepackage{xcolor}
\usepackage{lipsum}                     
\usepackage{xargs}                      
\usepackage[colorinlistoftodos,prependcaption,textsize=tiny]{todonotes}
\newcommand\concern[1]{#1}

\usepackage{varwidth}
\usepackage{enumitem}
\usepackage{mathtools, nccmath}

\usepackage[compact]{titlesec}
\usepackage{hyperref}
\setlist{noitemsep,topsep=0pt,parsep=0pt,partopsep=0pt,leftmargin=0.25cm}

\newcommandx{\unsure}[2][1=]{\todo[linecolor=red,backgroundcolor=red!25,bordercolor=red,#1]{#2}}
\newcommandx{\change}[2][1=]{\todo[linecolor=blue,backgroundcolor=blue!25,bordercolor=blue,#1]{#2}}
\newcommandx{\info}[2][1=]{\todo[linecolor=olive,backgroundcolor=olive!25,bordercolor=olive,#1]{#2}}
\newcommandx{\improvement}[2][1=]{\todo[linecolor=Plum,backgroundcolor=Plum!25,bordercolor=Plum,#1]{#2}}
\newcommandx{\thiswillnotshow}[2][1=]{\todo[disable,#1]{#2}}

\long\def\ignore#1{}

\newcommand\boldparagraph[1]{\noindent\textbf{#1}}

\title{Memory-efficient Speech Recognition on Smart Devices}

\name{\begin{tabular}{c}Ganesh Venkatesh, Alagappan Valliappan, Jay Mahadeokar, Yuan Shangguan, \\ Christian Fuegen, Michael L. Seltzer, Vikas Chandra\end{tabular}}

\address{Facebook Inc.}

\begin{document}
%
\maketitle
\begin{abstract}
Recurrent transducer models have emerged as a promising solution for speech recognition on the current and next generation smart devices. The transducer models provide competitive accuracy within a reasonable memory footprint  alleviating the memory capacity constraints in these devices. However, these models access parameters from off-chip memory for every input time step which adversely effects device battery life and limits their usability on low-power devices.

We address transducer model's memory access concerns by optimizing their model architecture and designing novel recurrent cell designs. We demonstrate that i) model's energy cost is dominated by accessing model weights from off-chip memory, ii) transducer model architecture is pivotal in determining the number of accesses to off-chip memory and just model size is not a good proxy, iii) our transducer model optimizations and novel recurrent cell reduces  off-chip memory accesses by \concern{$4.5\times$} and model size by \concern{$2\times$} with minimal accuracy impact.
\end{abstract}
\section{Introduction}
\label{sec:intro}

Speech is a natural interface for the ``smart'' devices around us -- especially the emerging class of keyboard-/screen-less devices such as assistant, watch, glass amongst others. Given the tremendous growth and adoption of these new devices, we expect speech to be primary mode of interaction for humans with their devices going forward. As a result, there is a lot of interest in building \emph{on-device} speech recognition to improve their reliability and latency as well as address user data privacy concerns. While the previous attempts at building on-device ASR involved scaling down traditional, memory-heavy multi-model system~\cite{speechalator,ctc} (acoustic/pronunciation/language models), recent work on Recurrent Transducer models~\cite{rnnt} (originally described in~\cite{rnntgraves}) have shown promise for the general problem of speech recognition using an end-to-end neural model. Their compact size addresses the memory capacity constraints while providing accuracy comparable to the larger server-side models.

\begin{figure}
 \centering
\includegraphics[width=0.9\columnwidth]{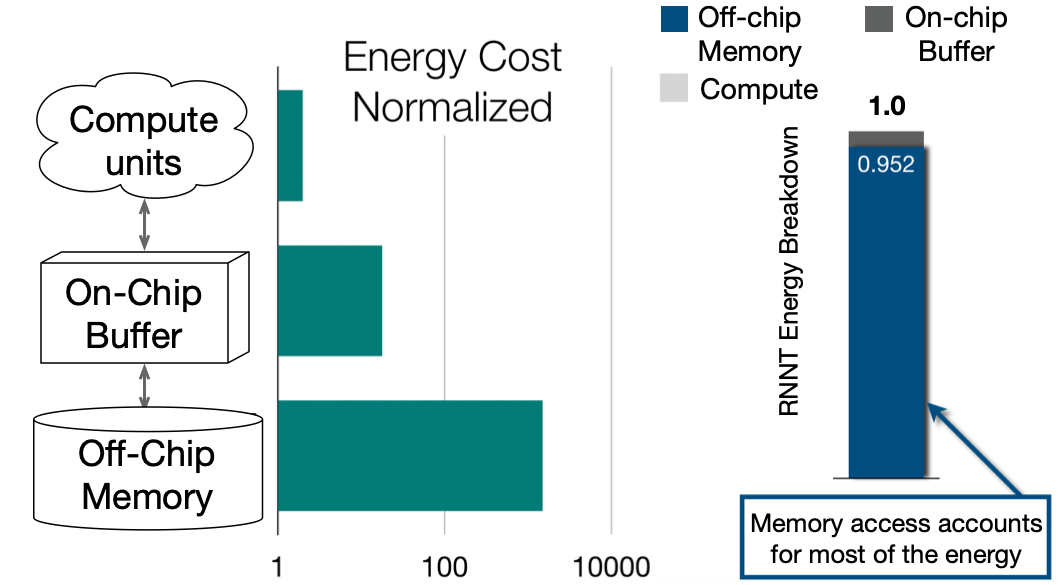}
 \caption{Hardware abstraction and energy cost breakdown~\cite{shakeynote,dlacclsurvey,horowitzenergy}. Transducer model power cost is dominated by its access to the off-chip memory to fetch weights.}
 \label{fig:rnntmem}
 \vskip -0.18in
\end{figure}

While the recurrent transducer models address the memory capacity constraints, their execution still relies on fetching weights from off-chip memory for every input speech frame. This repeated access of weights from off-chip memory makes the model inefficient because the cost of accessing memory is significantly higher than accessing on-chip buffers as well as performing computations (Figure~\ref{fig:rnntmem}). Transducer model's memory heavy behavior can limit their ability to run on low-end devices with slow off-chip memory. In this work, we significantly reduce off-chip memory accesses by redesigning transducer model such that it can access model parameters primarily from on-chip buffers. We find that the number of off-chip memory accesses is not just proportional to the model size but rather depends on the per-layer parameter count as well as how we schedule across time steps and layers. This work makes the following contributions:
\begin{itemize}
	\item \textbf{Efficient Transducer model architecture} that reduces the number of off-chip memory accesses for model parameters. 
	\item \textbf{Novel recurrent cell design} that models the cell state as a matrix instead of a vector. This gives us more flexibility in terms of sizing a layer to fit within on-chip buffers. 
	\item \textbf{Memory-efficient Model}: We reduce memory accesses by ~\concern{4.5$\times$} and model size by \concern{$2\times$} without loss in accuracy.
\end{itemize}

\section{Recurrent Transducer Model}
\label{sec:rnnt}

This section presents an overview of Recurrent Transducer networks \cite{rnnt,rnntgraves} and challenges in deploying them on low-end smart devices. 
The recurrent transducer network (Figure~\ref{fig:rnnt}) consists of three components -- \textbf{encoder} works on the input audio stream, \textbf{prediction}  uses the previously predicted symbols to guide the next symbol and \textbf{joint network} combines the two networks to produce the probability distribution of the next symbol. Encoder is the largest component of the transducer network and also executes the most often because the number of input speech frames are much higher than the output word pieces. We focus on  encoder in this work.

\subsection{Encoder Network Architecture}
\label{sec:encoder}

Encoder accepts as input the audio stream preprocessed into features and runs them through a multi-layer LSTM~\cite{lstm}  (Figure~\ref{fig:rnnt}) network to produce feature representation. Our encoder uses unidirectional LSTMs to target the streaming behavior.  

\noindent \textbf{Compute Building Block:} LSTMs take as input vector for
current time step ($x_t$) and hidden state from previous time step
($h_{t1}$). It uses three gates – input ($i$), forget ($f$) and output
($o$) to update the cell memory ($c$) and produce new hidden
state vector ($h_t$) for the next time step. Current day networks
typically stack multiple layers of the LSTM cells where the
hidden state output of a layer is fed as input to the next layer
as input ($x_{l+1}$). To aid with training stability and efficiency, 
speech models such as RNNT~\cite{rnnt} use LayerNorm~\cite{layernorm} (shown as $ln$ in equations) within the LSTM to normalize the cell, gate and output calculations.

\vspace*{2mm}
\begin{align*}
fp_t, ip_t, cp_t, o_t &= ln([W_f, W_i, W_C, W_o] \cdot [h_{t-1}, x_t]^T)\\
f_t, i_t, o_t &= sigmoid(fp_t, ip_t, op_t)\\
\tilde{c_t} &= tanh(cp_t) \\
c_t &= ln(f_t*c_{t-1} + i_t*\tilde{c_t})\\
h_t &= o_t*tanh(c_t)\\
\end{align*}
\vspace*{-8mm}

\noindent \textbf{Time Reduction Layer:} This layer merges features from neighboring time steps into a single embedding vector -- a common reduction technique being concatenation. This layer is motivated by the observation that there are many more input speech frames (one every 10 ms or so) than the number of output tokens (word pieces). Time reduction helps address this imbalance which improves the training stability~\cite{las}.

\subsection{Encoder Computation Scheduling}
\label{sec:encoder}

\boldparagraph{Execute one speech utterance per encoder inference:} The basic scheduling would run the whole encoder network for every new speech frame. In this scheme, we will fetch all the model parameters from off-chip memory every time step. 

\begin{figure}
 \centering
\includegraphics[width=\columnwidth]{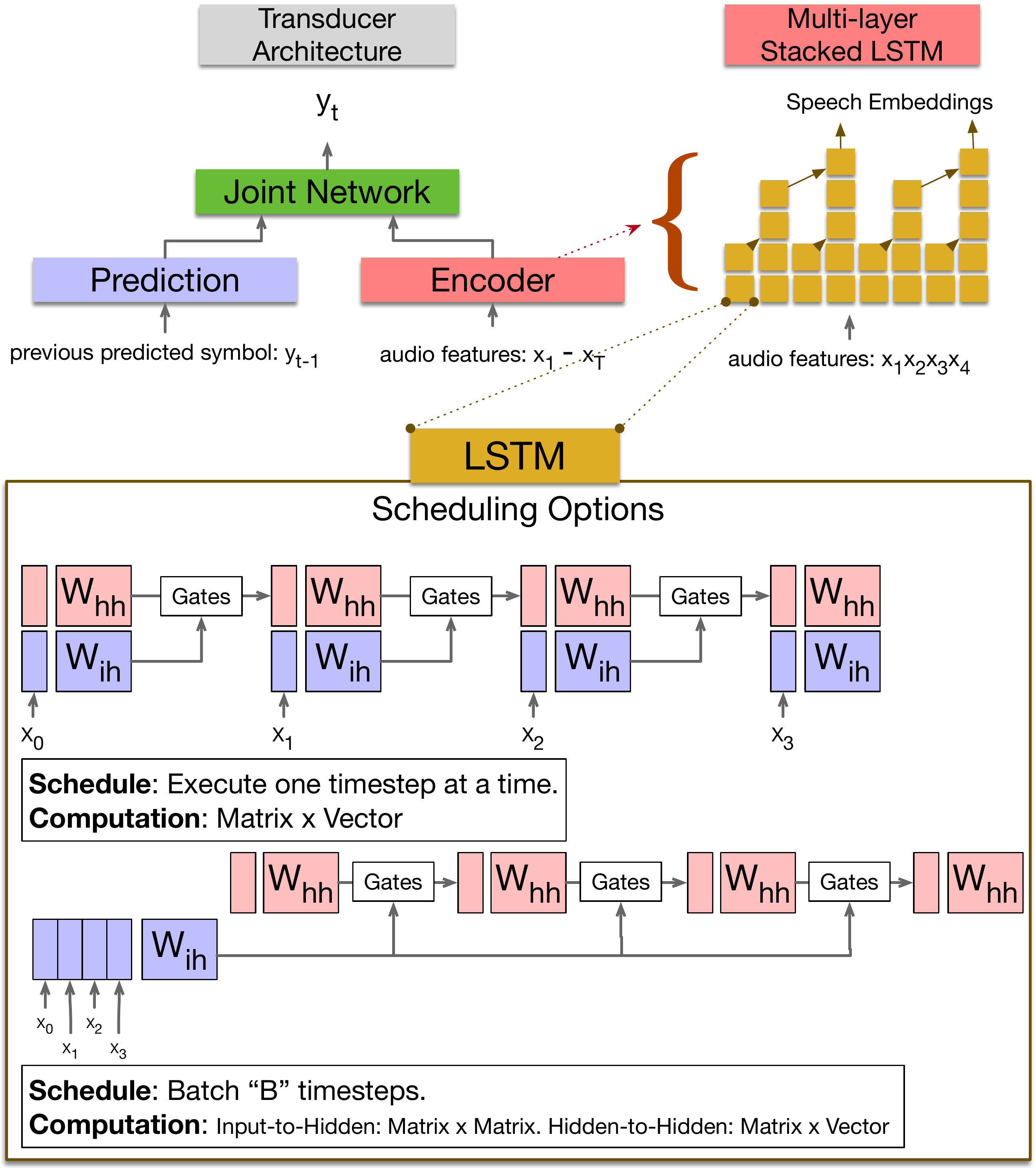}
 \caption{Recurrent Transducer Network}
 \vspace*{-3mm}
 \label{fig:rnnt}
\end{figure}

\boldparagraph{Batch ``B'' time steps:} An alternative would be to wait for \textbf{B} speech frames and execute encoder on all of them. By doing so, we fetch weights for the input-to-hidden path ($W_{ih}$ in Figure~\ref{fig:rnnt}) only once for ``B'' time steps. However, we still need to fetch weights for hidden-to-hidden path ($W_{hh}$ in Figure~\ref{fig:rnnt}) ``B'' times from off-chip memory because of the sequential dependence between each time step calculation. 

In either scheme, we need to fetch model parameters from off-chip memory for each input speech frame. Section~\ref{sec:lstmvar} discusses how to optimize away this need for repeated off-chip memory access of model parameters.

\subsection{Deployment challenges}
\label{lstmcons}

While LSTM-based encoder models achieve high accuracy for streaming speech recognition~\cite{rnnt}, there are many challenges in deploying them on smart devices.
\begin{description}
\item[Repeated access to model parameters:] These models need to access model parameters at each input time step as shown in Section~\ref{sec:encoder}. 

\item[Large memory footprint:] An LSTM layer has O($8d^2$) parameters each where $d$ is the dimension of the input/hidden state. As a result, for any reasonable layer dimension, it is too large to store a complete layer (or even subset such as $W_{hh}$) on-chip. This large memory footprint combined with the above observation about repeated access to model parameters result in frequent access to off-chip memory. This is the reason the model execution power is data access dominated as shown in Figure~\ref{fig:rnntmem}.
\item[Few optimization knobs:] The design space for these models is very coarse in that the size of a layer is controlled primarily by just one knob -- layer dimension. Furthermore, reducing layer dimension also limits the representation power -- cell memory and hidden state. 
\end{description}

\subsection{Efficient RNNT Variations} 
\label{sec:related}

Prior work has looked at addressing these RNN-T inefficiencies by using LSTM variants such as CIFG~\cite{lstmvar,rnntperfopt, sru}. While model size is a popular target metric for optimization, as we show in this work, it is not a good proxy for execution efficiency. Hence, we instead use expected memory access counts as our optimization metric. Our work focuses on developing novel LSTM variants and sizing them appropriately so that we can maintain speech model accuracy while improving its memory access efficiency. We demonstrate that there is scope for large, non-linear improvements in memory access efficiency by appropriately designing RNN-T model. 

In the next section, we explore transducer model variants that improve network's efficiency and flexibility.

\section{Optimizing Transducer Models}
\label{sec:lstmvar}

We inspect various components of the transducer encoder and propose variations to improve model efficiency. In particular, we optimize layer normalization, time reduction, simplify multi-layer stacking and allow reducing layer size without hurting its cell memory size. We show our modified cell equation and highlight the changes with a box.

\subsection{Layer Normalization}
\label{sec:ln}

Layer Normalization ($ln$) helps stabilize training by normalizing the output of the compute layers. The original implementation normalizes values across all the gates ($i$, $f$, $o$ gate) and cell state ($c$). Given the very different downstream usage of gates and cell state, it is not clear why they should be normalized together. An alternative would be normalizing them separately~\cite{lnind}, but that makes it hard for gates to be all small or large needed to mostly forget or carry over past state. 

We use layer normalization only to normalize cell state computation. In our experiments, we see that this provides similar training stability while being more efficient at inference time because we can skip normalization of gates. \useshortskip

\begin{align*}
\Aboxed{fp_t, ip_t, cp_t, o_t &= ([W_f, W_i, W_c, W_o] \cdot [h_{t-1}, x_t]^T)} \\
f_t, i_t, o_t &= sigmoid(fp_t, ip_t, op_t)\\
\Aboxed{\tilde{c_t} &= tanh(ln(cp_t))} \\
c_t &= ln(f_t*c_{t-1} + i_t*\tilde{c_t})\\
h_t &= o_t*tanh(c_t)\\
\end{align*}
\vspace{-0.5in}

\subsection{Internally Stacked Recurrent Cells}
\label{sec:dlstm}

This technique simplifies layer stacking and reduces its cost. The current approach for multi-layer stacked recurrent networks stack another LSTM layer on top. Our approach explores stacking internal to the recurrent cell by just building a deeper network for cell memory and hidden state computation. This provides a network designer additional knobs in terms of achieving extra depth with minimal increase in parameters and network complexity.\useshortskip  

\begin{align*}
fp_t, ip_t, cp_t, o_t &= ([W_f, W_i, W_c, W_o] \cdot [h_{t-1}, x_t]^T) \\
f_t, i_t, o_t &= sigmoid(fp_t, ip_t, op_t)\\
\Aboxed{\tilde{c_t} &= tanh(ln(W_{ch} \cdot cp_t))} \\
c_t &= ln(f_t*c_{t-1} + i_t*\tilde{c_t})\\
h_t &= o_t*tanh(c_t)\\
\end{align*}
\vspace{-0.5in}

\subsection{Two-dimensional Cell Memory}
\label{sec:veclstm}

We extend the traditional LSTM cells with two-dimensional cell memory instead of a vector which is single dimensional. The motivation is as follows -- when we reduce the LSTM layer size by reducing the hidden-state size, we also reduce the memory capacity of the layer since cell memory size and hidden state is proportional to the hidden size. In our approach, we model cell memory as $h\times v$ matrix where $h$ is the hidden state size and $v$ is the number of cell memory vectors in the recurrent cell. By doing so, we can reduce the hidden state size without reducing the cell memory by increasing $v$.\useshortskip

\begin{align*}
fp_t, ip_t, cp_t, o_t &= ([W_f, W_i, W_c, W_o] \cdot [h_{t-1}, x_t]^T) \\
f_t, i_t, o_t &= sigmoid(fp_t, ip_t, op_t)\\
\Aboxed{Cm_t &= ln(W_{ch} \cdot cp_t).view(h, v)} \\
\Aboxed{\tilde{C_t} &= tanh(Cm_t)} \\
\Aboxed{C_t &= ln(f_t*C_{t-1} + i_t*\tilde{C_t})}\\
\Aboxed{Hm_t &= o_t*tanh(C_t)}\\
\Aboxed{h_t &= Hm_t.view(h\times v)} \\
\end{align*}
\vspace{-0.5in}
\subsection{Time Reduction Layer}
\label{sec:label}

To reduce the number of tokens flowing through the encoder network, a common time reduction layer step is using concatenation~\cite{rnnt}. This does come at the cost of an increase in number of parameters. We simplify the time reduction step by replacing concatenation to mean. By doing so, we can i) reduce the model size without impacting accuracy, ii) reduce the number of memory accesses and compute in a pretrained model by changing number the time reduction factor without any change to network weight shape. This allows us to quickly adapt a trained model to smart device's memory constraints without having to train a new model.

\section{Results}
\label{sec:results}

In this section, we demonstrate that i) model size is a poor proxy to approximate it's expected off-chip memory accesses, ii) our transducer architecture optimizations and novel recurrent cell  reduces the encoder size by \concern{2.5$\times$} and iii) we achieve super-linear savings in off-chip memory access of \concern{4.5$\times$} demonstrating that model size is not proportional to off-chip memory access. To demonstrate the above points, we construct and train the following model variations (Table~\ref{tab:rnntacc}):

\begin{description}
\item[B] This is our baseline network based on RNNT paper~\cite{rnnt} that uses multi-layer stacked LSTMs. We scale the model size down to below \concern{40 MB}. This model's encoder has two time reductions layers that each reduce the token count by $2\times$.
\item[E1] We change time reduction operator from concatenation to mean with no impact on accuracy.
\item[E2] We build a deeper variant of the network where each layer is skinnier (\concern{1.78$\times$} smaller). However, it does not converge.
\item[E3] Replacing LSTM with its residual~\cite{residuallstm} variant helps the network converge and recover the original accuracy.
\item[E4] Replacing LSTMs with CIFG~\cite{lstmvar} helps us achieve further reduce model size for a very modest accuracy hit.
\item[E5] Adding greater depth via internal stacking (IS) following the time reduction layer helps us recover accuracy with almost no increase in parameter count.
\item[E6] Replacing CIFG with our novel two-dimensional cell memory (2D) allows us to reduce hidden dimension from 480 to 256 with minimal accuracy impact. Our novel design reduces the hidden dimension of the recurrent cell without reducing the cell memory size.
\item[E7] We further reduce the model size by building a deeper model with skinnier cells (200 hidden state). 
\end{description}

\subsection{Optimizing Off-chip Memory Accesses}
\label{sec:dataeff}

\begin{figure}
 \centering
\includegraphics[width=\columnwidth]{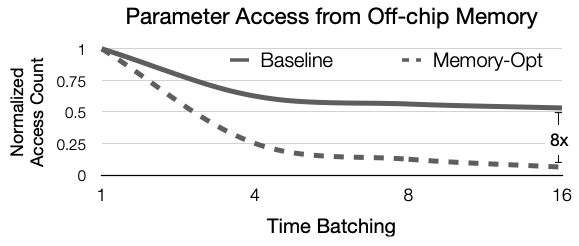}
\vspace*{-6mm}
 \caption{Reducing model parameter access from Off-chip memory by exploiting on-chip buffers (Memory-Opt)}
 \vspace*{-3mm}
 \label{fig:dataeff}
\end{figure}

We start by demonstrating that designing a model to be on-chip buffer aware is critical. For this discussion, we focus on one LSTM layer with hidden state size $H$ and say it can work on $T$ input samples at once (for streaming use cases T can be 4 - 16 incurring 100 - 200 ms model latency). As shown in Figure~\ref{fig:dataeff}, when a layer's recurrent path ($W_{hh}$ in Figure~\ref{fig:rnnt}) can fit on-chip (Memory-Opt), then it accesses off-chip memory much less often (up to \concern{8$\times$} lower) than the baseline scenario where the layer needs to fetch weights from off-chip memory for every speech utterance. Since off-chip memory accesses dominate Transducer's execution (Figure~\ref{fig:rnntmem}), replacing them with on-chip memory accesses enables efficient speech recognition on low-end devices.

To realize the above gains, each layer should be small enough to fit within on-chip buffers. In the next section, we show how our optimizations can reduce the per-layer size by more than \concern{3$\times$} with minimal accuracy impact. 

\subsection{Accuracy Evaluation}
\label{sec:rnntacc}

We train the model variations on Librispeech~\cite{librispeech} -- we use ADAM optimizer for 75 epochs with 61 epochs on a constant learning rate of 0.0004 and polynomial scaling by factor of 0.8. Table~\ref{tab:rnntacc} shows the results from the various configurations. We would like to highlight the following observations:
\begin{description}
\item[Significant Model Size Reductions] Our model and layer optimizations can reduce the model size by \concern{2$\times$} with minimal accuracy impact. Furthermore, we spread the smaller model size across a greater number of layers which reduces on average per-layer size by more than \concern{3$\times$}. This makes the design amenable to utilizing on-chip buffers for each layer.
\item[LayerNorm on Cell state is enough] Experiments from E5-E7 only use layer normalization for the cell state and we did not see any network training stability issues.
\item[Internal Stacking is Efficient (E4 $\rightarrow$ E5)] Using internal stacking to increase depth by 2 improves accuracy with negligible increase in the model size.
\item[2D cell state allows hidden-state reduction (E5 $\rightarrow$ E7)] We can reduce the hidden state dimension significantly (E6 and E7) with modest accuracy impact accuracy by representing the cell memory and hidden state as a 2D matrix. This allows us to reduce a layer's parameter count without impact its representation power.
\end{description}

\begin{table}[htbp]
\caption{Accuracy, Parameter Counts and Cell Variations}
\label{tab:rnntacc}
\vskip -0.02in
\centering
\resizebox{\columnwidth}{!}{%
\begin{tabular}{cccccccc}
\toprule
& & & & & \multicolumn{2}{c}{Param (M)} & \\
ID & Gate & H & Vec & Depth & Network & Encoder & WER \\
\midrule
B & LSTM & 640 & 1 & 8 & 37 & 32.8 & 4.63 \\
E1 & LSTM & 640 & 1 & 8 & 34 & 29.5 & 4.63 \\
E2 & LSTM & 480 & 1 & 11 & 28 & 23.6 & Diverges \\
E3 & LSTMR & 480 & 1 & 11 & 28 & 23.6 & 4.63 \\
E4 & CIFGR & 480 & 1 & 11 & 24 & 19.5 & 4.74 \\
E5 & IS-CIFGR & 480 & 1 & 11 & 24.5 & 20 & 4.63 \\
E6 & IS-2D-CIFGR & 256 & 2 & 11 & 22 & 17.5 & 4.89 \\
E7 & IS-2D-CIFGR & 200 & 2 & 12 & 18 & 13.2 & 4.87 \\
\bottomrule
\end{tabular}
}
\vskip -0.1in
\end{table}

\subsection{Efficiency Evaluation}
\label{sec:rnnteff}

This section discusses the improvement in execution efficiency from our model optimizations. For this analysis, we assume a mobile system with a modest resource of \concern{$\sim$512 KB}~\cite{cortexa77} on-chip buffer. Table~\ref{tab:rnntopt} shows reduction in off-chip memory accesses -- we reduce it to \concern{0.57$\times$} by optimizing the transducer model architecture (E3) and  to \concern{0.22$\times$} by using our novel recurrent cells. We also note that reduction in memory accesses are higher than the reduction in the model size reinforcing our observation that the savings come from a more efficient model design and not just from model size reduction.

\begin{table}[htbp]
\vskip -0.1in
\caption{Model Efficiency Improvements}
\label{tab:rnntopt}
\vskip -0.02in
\centering
\resizebox{\columnwidth}{!}{%
\begin{tabular}{ccccc}
\toprule
ID & Param (M) & WER (C) & Encoder Size & Off-chip Memory \\
\midrule
B & 37 & 4.63 & \concern{1$\times$} & \concern{1$\times$} \\
E3 & 28 & 4.63 & \concern{0.7$\times$} & \concern{0.57$\times$} \\
E7 & 18 & 4.87 & \concern{0.4$\times$} & \concern{0.22$\times$} \\
\bottomrule
\end{tabular}
}
\vskip -0.1in
\end{table}

\section{Discussion and Future Directions}
\label{sec:future}

Our work significantly reduces the memory traffic in a speech model inference without compromising on the accuracy. We believe this will be critical in providing high-quality speech support on the next generation devices such as smart-watch, AR-glass among others that will have limited memory resources as well as severe power constraints. Further efficiency gains are possible by combining our approach with pruning, quantization and neural architecture search. 
Another related line of work would be post-training adaptation of a trained speech recognition model for deployment on different target devices, since the memory traffic overhead on a phone, watch and glass will vary substantially. Our technique of using \textit{mean} in the time-reduction layer instead of \textit{concat} enables adapting a model post-training. For example, by increasing the time-reduction factor we can reduce the memory accesses by more than \concern{20\%} while recovering much of the accuracy with quick fine-tuning. This in combination with other techniques such as LayerDrop~\cite{layerdrop} can provide significant deployment flexibility and reduce the need to train many different models.
\section{Conclusion}
\label{sec:conclude}

We propose an optimized transducer model architecture built with a novel recurrent cell design that reduces its off-chip memory accesses by \concern{$4.5\times$} and model size by \concern{$2.5\times$}. With our architecture optimizations, we can enable high accuracy speech recognition support on low-end smart devices.

\vfill\pagebreak

\small
\bibliographystyle{IEEEbib}
\bibliography{main}

\end{document}